\documentclass[conference]{IEEEtran}
\IEEEoverridecommandlockouts
\usepackage{cite}
\usepackage{amsmath,amssymb,amsfonts}
\usepackage{algorithmic}

\usepackage{ifluatex}
\ifluatex
  \usepackage{pdftexcmds}
  \makeatletter
  \let\pdfstrcmp\pdf@strcmp
  \let\pdffilemoddate\pdf@filemoddate
  \makeatother
\fi

\usepackage{svg}
\usepackage{graphicx}
\usepackage{textcomp}
\usepackage{xcolor}
\usepackage{soul}
\def\BibTeX{{\rm B\kern-.05em{\sc i\kern-.025em b}\kern-.08em
    T\kern-.1667em\lower.7ex\hbox{E}\kern-.125emX}}
\newtheorem{definition}{Definition}
\usepackage[left=0.75in, right=0.75in, bottom=0.75in, top=1in]{geometry}
\newtheorem{theorem}{Theorem}
\usepackage{hyperref}

\begin{document}
% \pagenumbering{arabic}
% \title{Applications of Optimal Control and Control Barrier Functions to Economics}
% \title{Optimal Control and Control Barrier Functions for Economics}
% \title{Economics and Control Barrier Functions}
\title{Stochastic Control Barrier Functions for Economics}

\author{David E.J. van Wijk$^{1}$\thanks{$^{1}$Aerospace Engineering Department, Texas A\&M University, College Station, TX, 77845. \texttt{\{davidvanwijk\}}\texttt{@tamu.edu}.}}

\maketitle

\begin{abstract}
Control barrier functions (CBFs) and safety-critical control have seen a rapid increase in popularity in recent years, predominantly applied to systems in aerospace, robotics and neural network controllers. Control barrier functions can provide a computationally efficient method to monitor arbitrary primary controllers and enforce state constraints to ensure overall system safety. One area that has yet to take advantage of the benefits offered by CBFs is the field of finance and economics. This manuscript re-introduces three applications of traditional control to economics, and develops and implements CBFs for such problems. We consider the problem of optimal advertising for the deterministic and stochastic case and Merton's portfolio optimization problem. Numerical simulations are used to demonstrate the effectiveness of using traditional control solutions in tandem with CBFs and stochastic CBFs to solve such problems in the presence of state constraints.
\end{abstract}

\section{Introduction}

Safety of autonomous systems is an important area of study that has received much attention in recent years, with applications to autonomous driving, robotics, and aerospace systems. Generally these problems are posed as constrained control problems, where certain regions of the state-space are classified as unsafe and are therefore off limits. Classical control theory has approached such constrained problems using model predictive control (MPC) \cite{muske1993model}, reachability analysis \cite{bansal2017hamiltonjacobi}, and reference governors \cite{governor2017}. 

Currently, a popular approach to assuring safety is the use of control barrier functions (CBFs) \cite{ames2019control}. Control barrier functions are continuous functions that decrease to zero at the boundary of safe regions. These functions can be used to derive conditions on the control to guarantee forward invariance of this safe set, while avoiding the complex computation of similar approaches such as reachable sets. CBFs can be applied to control systems, where optimization-based methods are often used to modify a desired control output in a minimally invasive manner while assuring safety.

A class of dynamical systems that has yet to be investigated for safety-critical control is that of financial and economic systems. However, there is a rich history of applying traditional control approaches to such systems. Following the rapid growth of the field of optimal control used in aerospace applications, economists became interested in applying this theory to macroeconomics. A survey paper from the 1970's collected around 60 papers pertaining to the application of control theory to macroeconomics \cite{kendrick_applications_1976}. These applications use what is now referred to as traditional control methods such as proportional–integral–derivative (PID) control or optimal control. Phillips compared the performance of proportional, proportional-integral, and PID control to a market stabilization policy in a closed economy \cite{phillips_stabilisation_1954}. Dobell and Ho posed the problem of investment of resources as an optimal control problem, and solved it using indirect methods \cite{dobell_optimal_1967}. During a similar time period, Sethi and Thompson formulated several versions of dynamic cash balance problems and used optimal control to solve them, while providing insight to financial interpretations of the Hamiltonian and the costates in these optimal solutions \cite{sethi_applications_1970}. Following the explosion of optimal control problems applied to finance and economics, several textbooks were developed on the subject \cite{sethi_optimal_2018,weber_optimal_2011}. 

As the hype of deterministic solutions began to subside, economists and financial analysts began to transition their efforts to modeling systems with stochasticity, more closely capturing the evolution of real world variables of interest. Sethi explored stochastic dynamic advertising using stochastic optimal control and compared it with deterministic models \cite{sethi_deterministic_1982}. Fleming and Pang applied stochastic control theory to a portfolio optimization problem, seeking to maximize expected discounted risk aversion \cite{fleming_application_2004}. Noh and Kim expanded the application in \cite{fleming_application_2004}, and considered stochastic volatility as well as stochastic interest in developing an optimal portfolio selection solution using the Hamilton–Jacobi–Bellman equations \cite{noh_optimal_2011}. Hudgins and Na developed H-infinity robust control solutions to financial asset and purchasing decisions in the presence of disturbances \cite{hudgins_h-optimal_2013}. The authors demonstrated the effectiveness of the approach for two examples using numerical simulations. Again, expansive textbooks on the subject followed \cite{cartea_algorithmic_2015,pham_continuous-time_2009}.

There has not been an attempt to apply recent advances in safety-critical control to economics applications posed as control problems. As such, the central contributions of this study are two-fold. Firstly, control barrier functions for a deterministic optimal advertising problem, a stochastic optimal advertising problem and for a portfolio optimization problem are introduced and justified. Secondly, the effectiveness of these approaches at satisfying safety constraints while achieving performance objectives is demonstrated using numerical simulations. 

The remainder of this study is organized as follows: First, background information on classical CBFs and safe control methods for nonlinear affine systems will be introduced. Second, recent advances in stochastic CBFs applied to stochastic systems will be summarized. Third, the three economics applications are presented and a safe feedback controller is derived for each by combining CBFs and optimal control. Fourth, the numerical results are summarized. Lastly, we conclude and indicate future directions for the work.

% \newpage

\section{Preliminaries} \label{sec:prelim}

This section introduces the fundamental concepts for CBFs and common extensions.

\subsection{Control Barrier Functions}\label{sec:cbf}

First, consider a control affine dynamical system modeled as
\begin{align} \label{eq:affine-dynamics}
    \dot{\boldsymbol{x}} = f(\boldsymbol{x}) + g(\boldsymbol{x})\boldsymbol{u},
\end{align}
where $f(\boldsymbol{x}):\mathcal{X} \rightarrow \mathbb{R}^n$ and $g(\boldsymbol{x}):\mathcal{X} \rightarrow \mathbb{R}^{n \times m}$ are Lipschitz continuous functions, $\boldsymbol{x} \in \mathcal{X} \subseteq \mathbb{R}^n$ represents the state vector, and $\boldsymbol{u} \in \mathcal{U} \subseteq \mathbb{R}^m$ represents the control vector. Safety of this system is enforced through a forward invariant set $\mathcal{C}_S$, known as the safe set.

\begin{definition}
A set $\mathcal{C} \subset \mathbb{R}^n$ is forward invariant for a dynamical system defined in Eq. \eqref{eq:affine-dynamics} when $\boldsymbol{x}(0) \in \mathcal{C}$ and $\boldsymbol{x}(t) \in \mathcal{C}, \, \forall t > 0$.
\end{definition}

Assume that the set $\mathcal{C}_S \subset \mathbb{R}^n$ is defined by a continuously differentiable function $h : \mathcal{X} \rightarrow \mathbb{R}$ where
\begin{align} \label{eq:safeset}
    \mathcal{C}_S = \{\boldsymbol{x} \in \mathcal{X} : h(\boldsymbol{x}) \ge 0\}, \\ \partial\mathcal{C}_S = \{ \boldsymbol{x} \in \mathcal{X} : h(\boldsymbol{x})=0 \}, \\
    \text{Int}(\mathcal{C}) = \{\boldsymbol{x} \in \mathcal{X} : h(\boldsymbol{x}) > 0 \}. \label{eq:safeset2}
\end{align}
\noindent Nagumo's condition \cite{nagumo1942lage} gives necessary and sufficient conditions for set invariance by ensuring $\dot{h}(\boldsymbol{x}) \geq 0$ along the boundary of $\mathcal{C}_S$. This ensures that $\boldsymbol{x}$ will never leave $\mathcal{C}_S$ as long as it starts in $\mathcal{C}_S$. Since this condition is only required at the boundary of $\mathcal{C}_S$, an extended class $\mathcal{K}$ function $\alpha$ is used to relax the constraint away from the boundary. 

\begin{definition} \label{def:2} 
A continuous function $\alpha : \mathbb{R} \rightarrow \mathbb{R}$ is an extended class $\mathcal{K}$ function if it is strictly increasing and has the property $\alpha(0) = 0$.
\end{definition}

\begin{definition}
Given a set $\mathcal{C}_S$ defined by Eqs.\ref{eq:safeset}$-$\ref{eq:safeset2} a function $h : \mathcal{X} \rightarrow \mathbb{R}$ is a zeroing control barrier function (ZCBF)\cite{amesOG} if there exists an extended class $\mathcal{K}$ function $\alpha$ such that
\begin{equation}\label{eq:cbf_condition1}
    \sup_{\boldsymbol{u} \in \mathcal{U}} [\dot{h}(\boldsymbol{x}, \boldsymbol{u}) + \alpha(h(\boldsymbol{x})) ] \geq 0, \forall \boldsymbol{x} \in \mathcal{C}_S.
\end{equation}
\end{definition}

% $h$ is then a Control Barrier Function (CBF)\cite{ames2019control} if,
% \begin{equation}
%     \exists \, \boldsymbol{u} \,\, \text{s.t.} \,\, \dot{h}(\boldsymbol{x}, \boldsymbol{u}) + \alpha(h(\boldsymbol{x})) \geq 0 \Leftrightarrow \mathcal{C}_S \,\, \text{is invariant}
% \end{equation}

\noindent For the remainder of the manuscript, it is assumed that all CBFs will be zeroing CBFs and thus will simply be called CBFs. For the control affine system defined in Eq. \eqref{eq:affine-dynamics}, the safety condition becomes
\begin{equation} \label{eq:BC}
       \sup_{\boldsymbol{u} \in \mathcal{U}} [\nabla h^{T}(\boldsymbol{x}) (f(\boldsymbol{x}) + g(\boldsymbol{x})\boldsymbol{u}) + \alpha(h(\boldsymbol{x}))] \ge 0, \forall \boldsymbol{x} \in \mathcal{C}_S.
\end{equation}
In another form, this is
\begin{equation} \label{eq:BC2}
   \sup_{\boldsymbol{u} \in \mathcal{U}} [L_f h(\boldsymbol{x}) + L_g h (\boldsymbol{x}) \boldsymbol{u} + \alpha(h(\boldsymbol{x}))] \geq 0
\end{equation}
for all $\boldsymbol{x} \in \mathcal{C}_S$, where $L_f$ and $L_g$ are Lie derivatives of $h$ along $f$ and $g$ respectively.

\begin{figure} [ht!]
    \centering
    \vspace{-.2cm}
    \centerline{\includegraphics[width=1\columnwidth]{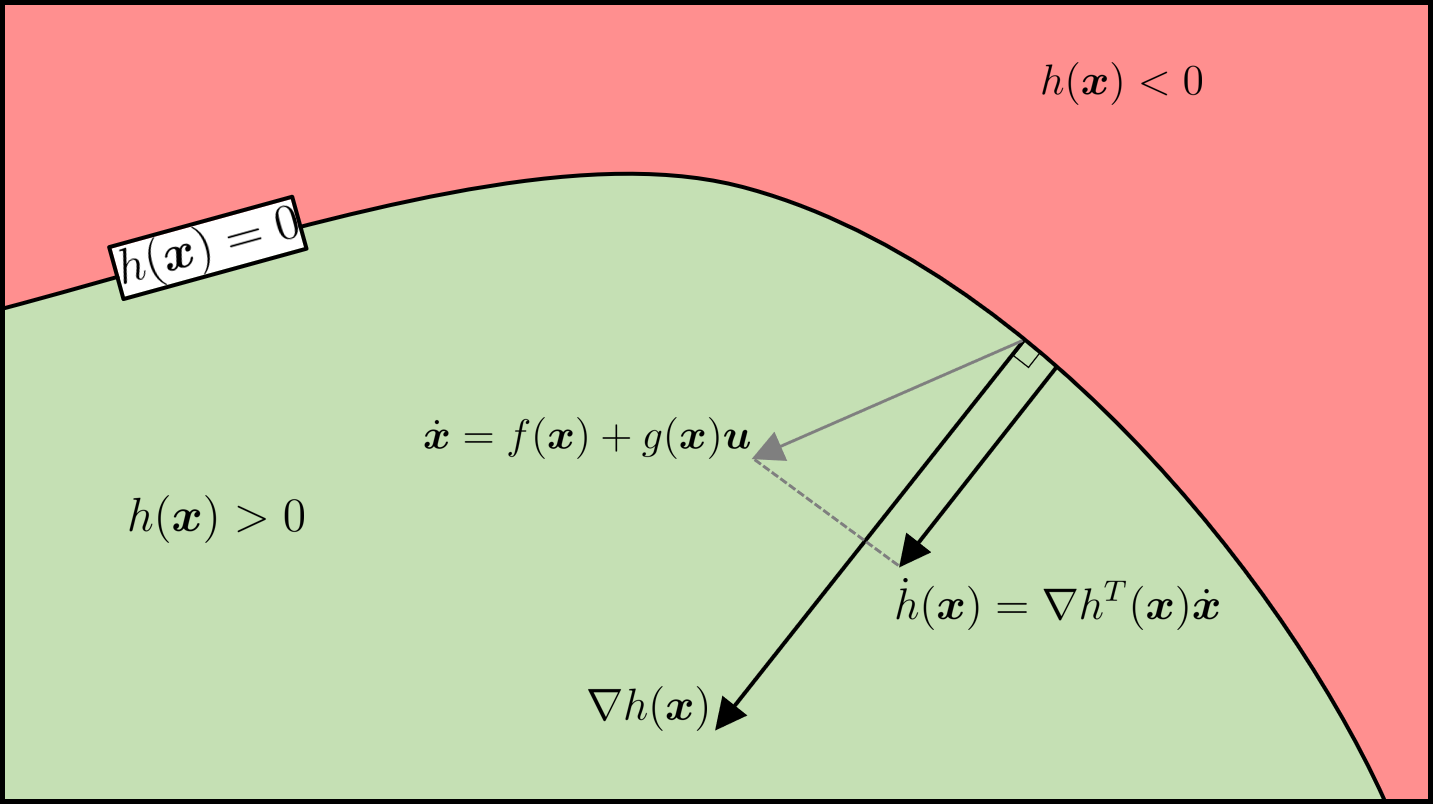}}
    \caption{Visualization of safety condition ensuring forward invariance of $\mathcal{C}_S$ without strengthening function.}
    \label{fig:cbf_boundary}
\end{figure}

\noindent Figure \ref{fig:cbf_boundary} provides a visual representation of the condition that must be satisfied at the boundary of $\mathcal{C}_S$. The red region is the unsafe region of the state-space while the green region is the safe region that the dynamical system must be kept within. With the figure, it is straightforward to understand what the condition is doing. It is forcing the dynamical system into the safe region when the system is at the boundary of the unsafe region, at $h(\boldsymbol{x})=0$. When a strengthening function is introduced, the condition is enforced before the dynamical system reaches the boundary.

\subsection{Active Set Invariance Filter}

One method of assuring safety of the system is with an active set invariance filter (ASIF), which is a first-order optimization-based algorithm designed to be minimally invasive with respect to safety constraints \cite{ASIF_2018}. Typically, ASIF uses a quadratic program (QP) to minimize the $l^2$ norm difference between the desired control input from some primary controller $\boldsymbol{u}_{\rm des}$ and actual control input $\boldsymbol{u}_{\rm act}$. The condition in Eq. \eqref{eq:BC2} is used as an inequality constraint to the QP, referred to as a barrier constraint, $BC(\boldsymbol{x}, \boldsymbol{u})\geq0$. The pointwise ASIF controller is defined as

\noindent \rule{1\columnwidth}{0.7pt}
\noindent \textbf{Active Set Invariance Filter}
\begin{equation}
\begin{gathered}
\boldsymbol{u}_{\rm act}(\boldsymbol{x}, \boldsymbol{u}_{\rm des})= \underset{\boldsymbol{u} \in \mathcal{U}}{\text{argmin}} \left\Vert \boldsymbol{u}_{\rm des}-\boldsymbol{u}\right\Vert,\\
\text{s.t.} \quad L_f h(\boldsymbol{x}) + L_g h (\boldsymbol{x}) \boldsymbol{u} + \alpha(h(\boldsymbol{x})) \geq 0.
\end{gathered}
\end{equation}
\noindent \rule[7pt]{1\columnwidth}{0.7pt}

% \noindent \rule{1\columnwidth}{0.7pt}
% \noindent \textbf{Active Set Invariance Filter}
% \begin{equation*}
% \begin{gathered}
% \boldsymbol{u}_{\rm act}(\boldsymbol{x}, \boldsymbol{u}_{\rm des})= \underset{\boldsymbol{u} \in \mathcal{U}}{\text{argmin}} \left\Vert \boldsymbol{u}_{\rm des}-\boldsymbol{u}\right\Vert ^{2}\\
% \text{s.t.} \quad L_f h(\boldsymbol{x}) + L_g h (\boldsymbol{x}) \boldsymbol{u} + \alpha(h(\boldsymbol{x})) \geq 0,
% \end{gathered}
% \end{equation*}
% \noindent \rule[7pt]{1\columnwidth}{0.7pt}

\noindent This ensures that the control input for the dynamical system, $\boldsymbol{u}_{\rm act}$, will guarantee safety and will be as close as possible to the control generated by the primary controller. An active set invariance filter using a QP is denoted ASIF-QP. In the literature similar formulations are often denoted CBF-QP's.

\subsection{Stochastic Control Barrier Functions}

With the groundwork for deterministic solutions laid, we direct our attention to safety assurance in the presence of stochasticity, and introduce stochastic CBFs (SCBF) \cite{prajna_2007,clark_control_2020}. Consider the dynamical system described by the stochastic differential equation (SDE)
\begin{align} \label{eq:SDE}
    {\rm d}\boldsymbol{x}(t) = (f(\boldsymbol{x}(t)) + {g}(\boldsymbol{x}(t))\boldsymbol{u}(t)){\rm d}t + \sigma(\boldsymbol{x}(t)){\rm d}\boldsymbol{w}(t),
\end{align}
where $f(\boldsymbol{x}(t)):\mathcal{X} \rightarrow \mathbb{R}^n$, $g(\boldsymbol{x}(t)):\mathcal{X} \rightarrow \mathbb{R}^{n \times m}$ and $\sigma(\boldsymbol{x}(t)):\mathbb{R}^n\rightarrow\mathbb{R}^q$ are Lipschitz continuous functions, $\boldsymbol{w}(t)$ is a Brownian motion, $\boldsymbol{x} \in \mathcal{X} \subseteq \mathbb{R}^n$ represents the state vector, and $\boldsymbol{u} \in \mathcal{U} \subseteq \mathbb{R}^m$ represents the control vector.  

The safe set, $\mathcal{C}_S$, is still defined by a continuously differentiable function, $h : \mathbb{R}^n \rightarrow \mathbb{R}$ satisfying Eq.~\ref{eq:safeset} and Eq.~\ref{eq:safeset2}. As with Sec.~\ref{sec:cbf}, the goal is to ensure forward invariance of the safe set, $\mathcal{C}_S$, thereby ensuring the safety of the dynamical system. However, the condition in Eq.~\ref{eq:BC2} is no longer sufficient. Due to the stochasticity of the dynamics in Eq.~\ref{eq:SDE}, the total derivative of $h(\boldsymbol{x})$ must be computed differently. 

For this, It$\hat{\text{o}}$'s Lemma \cite{oksendal_stochastic_2003} is used. If a twice differentiable function $h=h(t,\boldsymbol{x})$ and $\boldsymbol{x}(t)$ is a general diffusion satisfying
\begin{align}
    {\rm d}\boldsymbol{x}(t) = \nu(t,\boldsymbol{x}) {\rm d}t + \sigma(t,\boldsymbol{x}) {\rm d}\boldsymbol{w}(t),
\end{align}
where $\boldsymbol{w}$ is a $d_b$ Brownian motion, $\nu(t,\boldsymbol{x})$ is $d_x \times 1$, and $\sigma(t,\boldsymbol{x})$ is $d_x \times d_b$, it can be shown through a Taylor series expansion that the SDE of $h(t,\boldsymbol{x})$ is given by
% \begin{align}
%     {\rm d}h(t,\boldsymbol{x}) = (\frac{\partial h}{\partial t} + \frac{\partial h}{\partial \boldsymbol{x}}\nu + \frac{1}{2}\texttt{tr}\{ \frac{\partial h^2}{\partial \boldsymbol{x}^2}\sigma \sigma^T \}){\rm d}t + \\ (\frac{\partial h}{\partial \boldsymbol{x}}\sigma){\rm d}\boldsymbol{w}(t)
% \end{align}
\begin{align}
    \begin{split}
    {\rm d}h(t,\boldsymbol{x}) = (\frac{\partial h}{\partial t} + \frac{\partial h}{\partial \boldsymbol{x}}\nu + \frac{1}{2}\texttt{tr}\{\sigma^T \frac{\partial^2 h}{\partial \boldsymbol{x}^2} \sigma \}){\rm d}t \\ 
    + (\frac{\partial h}{\partial \boldsymbol{x}}\sigma){\rm d}\boldsymbol{w}(t). 
    % \triplequad \quad \quad
    \end{split}
\end{align}
% Here, $\texttt{tr}\{ \cdot \}$ represents the trace operator. Clark \cite{clark_control_2020} claimed that for a time-invariant control barrier function $h(\boldsymbol{x})$ and under the assumption that the barrier function is a \textit{supermartingale} then $\texttt{Pr}\{ \boldsymbol{x}(t) \in \mathcal{C}_S ~ \forall t\} = 1$ if $\boldsymbol{u}(t)$ satisfies
Here, $\texttt{tr}\{ \cdot \}$ represents the trace operator. Clark \cite{clark_control_2020} claimed that for a time-invariant control barrier function, $h(\boldsymbol{x})$, $\texttt{Pr}\{ \boldsymbol{x}(t) \in \mathcal{C}_S ~ \forall t\} = 1$ if $\boldsymbol{x}_0 \in \mathcal{C}_S$ and $\boldsymbol{u}(t)$ satisfies
\begin{align}
\begin{split} \label{eq:scbf_bc}
        \frac{\partial h}{\partial \boldsymbol{x}}(f(\boldsymbol{x}) + {g}(\boldsymbol{x})\boldsymbol{u}) + \frac{1}{2}\texttt{tr}\{\sigma(\boldsymbol{x})^{T} \frac{\partial^2 h}{ \partial \boldsymbol{x}^2} \sigma(\boldsymbol{x}) \} \\
    \ge -\alpha(h(\boldsymbol{x})),
\end{split}
\end{align}
for the dynamical system in Eq.~\ref{eq:SDE} for all $t$. Here $\texttt{Pr}\{ \cdot \}$ returns the probability of $\{ \cdot \}$. 
% A supermartingale is an integer-time stochastic process where the expected value evolution of the process is non-positive. This means that the expected value of any term in the process is either smaller or equal to the previous term.
This condition was used for many stochastic CBF applications in recent years, with over 50 citations.

However, in a paper published on December 5, 2023, So et al. demonstrate that the proof for the above formulation is flawed despite being widely used and well-cited. Thus Eq.~\ref{eq:scbf_bc} is not sufficient to guarantee safety in general \cite{so2023almostsure}. They use a simple counterexample of uncontrolled Brownian motion to contradict the proof, and then provide additional terms that must be included to assure safety of Eq.~\ref{eq:SDE}. The authors start by reexamining the SDE of $h$ for the system \eqref{eq:SDE} using It$\hat{\text{o}}$'s Lemma, with the assumption that $h$ does not depend on time
\begin{align}
    \begin{split}
    {\rm d}h(\boldsymbol{x}) = \underbrace{(\frac{\partial h}{\partial \boldsymbol{x}}(f(\boldsymbol{x}) + {g}(\boldsymbol{x})\boldsymbol{u}) + \frac{1}{2}\texttt{tr}\{\sigma(\boldsymbol{x})^{T} \frac{\partial^2 h}{ \partial \boldsymbol{x}^2} \sigma(\boldsymbol{x}) \} )}_{:=\Tilde{\mu}}{\rm d}t \\ 
    + \underbrace{(\frac{\partial h}{\partial \boldsymbol{x}}\sigma(\boldsymbol{x}))}_{:=\Tilde{\sigma}}{\rm d}\boldsymbol{w}, \qquad \qquad \qquad \qquad \qquad \qquad \qquad
    \end{split}
\end{align}
% \vspace{-.3cm}
\noindent where here $\Tilde{\mu}$ and $\Tilde{\sigma}$ represent the drift and diffusion terms of $h(\boldsymbol{x})$ respectively. 
% Using the drift and diffusion terms, the new condition for ensuring system safety is with
% \begin{align}
%     \Tilde{\mu} - \frac{\Tilde{\sigma}^2}{h(\boldsymbol{x})} \ge -h(\boldsymbol{x})^2\alpha_2(h(\boldsymbol{x}))
% \end{align}
% where $\alpha_2$ is an extended class $\kappa_\infty$ function.
Using the drift and diffusion terms we arrive at the main SCBF theorem.
\begin{theorem} \label{thm:1}
Suppose there exists a function $h : \mathcal{X} \rightarrow \mathbb{R}$ and a extended class $\mathcal{K}$ function $\alpha$ where, for all $x \in \mathcal{X}$ satisfying $h(\boldsymbol{x}) > 0$, there exists $\boldsymbol{u}\in \mathcal{U}$ such that 
\begin{align} \label{eq:scbf_BC_NEW}
    \Tilde{\mu} - \frac{\Tilde{\sigma}^2}{h(\boldsymbol{x})} \ge -h(\boldsymbol{x})^2\alpha(h(\boldsymbol{x})).
\end{align}
Then, for all $t \ge 0$, $\texttt{Pr}\{\boldsymbol{x}(t) \in C_S\}=1$, provided that $\boldsymbol{x}(0)\in \rm Int(\mathcal{C}_S)$.
\end{theorem}

Thus, to enforce safety for Eq.~\ref{eq:SDE}, one must now enforce Eq.~\ref{eq:scbf_BC_NEW} using an ASIF.

\section{Problem Formulation} \label{sec:problem_formulation}

Three problems from the economics and finance literature are chosen to motivate the use of CBFs and safety-critical control in a setting other than cyber-physical systems (CPS). The first problem is a one-dimensional problem of optimal advertising with a constraint on the market share. The second problem is a stochastic optimal advertising problem with a similar constraint. The third application is Merton's portfolio optimization problem with a risky asset. 

\subsection{Optimal Advertising} \label{sec:oa}

The optimal advertising problem introduced in \cite{weber_optimal_2011} considers the problem of maximizing the discounted profits of a firm using advertising.
% The advertising model was first introduced by Vidale and Wolfe \cite{vidale}.
The state of interest, $\boldsymbol{x} \in [0, 1]$, is the market share, which is the percentage of the population of potential customers who have already purchased the firm's product. The state evolves as
\begin{align} \label{eq:dynamics1}
    \dot{\boldsymbol{x}}(t) = (1-\boldsymbol{x}(t))a(t)^{\kappa} - \beta \boldsymbol{x}(t),
\end{align}

\noindent where $\beta > 0$ and $\kappa \in (0,1)$ are constants. The control variable is $a \in [0,a_{\rm max}]$ where $a_{\rm max} > 0$. The control variable reflects the intensity of the firm's advertising activity. The objective function is
\begin{align} \label{eq:obj_fun_1}
    J(a) = \int_{0}^{\infty} e^{-rt}((1-\boldsymbol{x}(t))a(t)^{\kappa} - c a(t)) dt,
\end{align}

\noindent which maximizes the firm's discounted profits where $c>0$ is the cost of advertising, given an initial state, $\boldsymbol{x}(0)=\boldsymbol{x}_0 \in (0,1)$. $r>0$ is a constant. The consumer demand equals the positive inflow to the market share, which is described by $(1-\boldsymbol{x})a(t)^{\kappa}$, and the firm is assumed to be a price taker in a durable goods market, meaning that the firm cannot dictate the prices in the market. In this application, the price has been normalized to one, and the products have a characteristic lifetime of $1/\beta$ before the goods must be disposed of. The constant $\kappa$ models the effect of diminishing return on advertising investments.

An additional assumption is added to the problem in the form of a state constraint. Assume that the firm has been accused of violating anti-trust laws in recent years and as such seeks to limit the market share to some percentage, $x_{\rm max} \in [0,1)$, to avoid breaking monopoly regulations. Therefore the new problem is to find the optimal control that maximizes Eq.~\ref{eq:obj_fun_1}, and is also monitored by an active set invariance filter equipped with a QP to efficiently intervene if the market share will exceed $x_{\rm max}$.

\subsubsection{Primary Controller}

An application of integration by parts using Eq.~\ref{eq:dynamics1}, readily yields $J(a)=c\hat{J}(\boldsymbol{u})-\boldsymbol{x}_0$, where $\boldsymbol{u}=a^{\kappa}$ and 
\begin{align}
    \hat{J}(\boldsymbol{u}) = \int_{0}^{\infty} e^{-rt}(\gamma \boldsymbol{x}(t) - \boldsymbol{u}^{\frac{1}{\kappa}}(t)) dt,
\end{align}

\noindent with $\gamma = \frac{r+\beta}{c}$. The control variable now becomes $\boldsymbol{u}$, and the primary controller is derived via indirect optimal control. The derivation starts from Pontryagin's maximum principle \cite{PMP}, with the Hamiltonian given by 
\begin{align}
    H(\boldsymbol{x}, \boldsymbol{u}, \boldsymbol{\lambda}, t) = e^{-rt}(\gamma \boldsymbol{x} - \boldsymbol{u}^{\frac{1}{\kappa}}) + \boldsymbol{\lambda} \dot{\boldsymbol{x}},
\end{align}

\noindent where $\boldsymbol{\lambda}$ denotes the costate, or the adjoint variable. Satisfying the necessary conditions for optimality
% \begin{subequations}
%     \begin{equation}
%         \frac{\partial H}{\partial \boldsymbol{u}} = 0,
%     \end{equation}
%     \begin{equation}
%         \frac{\partial H}{\partial \boldsymbol{x}} = -\dot{\boldsymbol{\lambda}},
%     \end{equation}
% \end{subequations}
\begin{align}
    \frac{\partial H}{\partial \boldsymbol{u}} = 0, \quad \frac{\partial H}{\partial \boldsymbol{x}} = -\dot{\boldsymbol{\lambda}}, \quad \frac{\partial H}{\partial \boldsymbol{\lambda}} = \dot{\boldsymbol{x}},
\end{align}
the optimal control solution is described by
\begin{subequations}
    \begin{equation}
        \boldsymbol{u}^{*}= 
        \begin{bmatrix}
        \frac{\kappa \boldsymbol{\lambda} (1 - \boldsymbol{x})}{e^{-rt}}
        \end{bmatrix}^{\frac{\kappa}{1-\kappa}},
    \end{equation}
    \begin{equation}
        \dot{\boldsymbol{\lambda}} = -\gamma e^{-rt} + \boldsymbol{\lambda}(\boldsymbol{u}^{*} + \beta).
    \end{equation}
\end{subequations}

\noindent Recalling the form of the ASIF algorithm, the optimal control is the desired control, $\boldsymbol{u}_{\rm des} = \boldsymbol{u}^{*}$. 

\subsubsection{Barrier Constraints}

The state constraint is straightforward in this application. The safety constraint function is formulated as
\begin{align}\label{eq:hoa}
    h_{oa}(\boldsymbol{x}) = -\boldsymbol{x}^2 + x_{\rm max}^2 \ge 0,
\end{align}

\noindent and therefore using the dynamics given in Eq.~\ref{eq:dynamics1}, the barrier constraint to the QP is
\begin{align}\label{eq:BC_advert}
    BC_{oa}(\boldsymbol{x}, \boldsymbol{u}) = 2\beta \boldsymbol{x}^2 - 2\boldsymbol{x}(1-\boldsymbol{x})\boldsymbol{u} + \alpha(h_{oa}(\boldsymbol{x})) \ge 0,
\end{align}
where $\alpha$ is a strengthening function as described in Sec.~\ref{sec:cbf}.

\subsection{Stochastic Optimal Advertising} \label{sec:soa}

The stochastic control example used is another example of optimal advertising, but with slight model modifications, as well as stochasticity injected into the system. The problem is a variation of the Vidale--Wolfe advertising model \cite{vidale} and was first introduced in \cite{sethi_deterministic_1982}, and is also featured in Ch.12 of \cite{sethi_optimal_2018}. In this example, the goal is still to maximize discounted profits, but since stochasticity is introduced, the \textit{expected value} of the discounted profits must be maximized. The market share, $\boldsymbol{x} \in [0,1]$ is the percentage of the population of potential customers who have already purchased the firm’s product. As with the deterministic case, this is the state of interest. However, it is now governed by an SDE given by
\begin{align} \label{eq:sde_oa1}
\begin{split}
    {\rm d} \boldsymbol{x}(t) = (-\beta \boldsymbol{x}(t) + r \sqrt{1-\boldsymbol{x}(t)}\boldsymbol{u}(t))\rm d t \\ 
    + \sigma(\boldsymbol{x})\rm d \boldsymbol{w}(t),
\end{split}
\end{align}
where $\boldsymbol{u}(t)$ is the control variable, representing the rate of advertising effort. $\boldsymbol{w}(t)$ is a one-dimensional Brownian motion and $\sigma(\boldsymbol{x})$ is the diffusion coefficient function. $\beta > 0$ is a sales decay parameter which accounts for loss of market share due to customer forgetting, product obsolescence, or competing advertising. The parameter $r > 0$ is the advertising effectiveness. The diffusion coefficient function in this example is given by $\sigma(\boldsymbol{x})=\sigma_a \boldsymbol{x}(t)$ where $\sigma_a$ is a constant standard deviation term. 

The problem is one of finding the optimal advertising rate which will maximize the expected value of discounted profits given by the objective function
\begin{align} \label{eq:obj3}
    J(\boldsymbol{u}) = \texttt{E}\{ \int^{\infty}_0 e^{-\rho t}(\pi \boldsymbol{x}(t) - \boldsymbol{u}^2(t))dt\}.
\end{align}

\noindent Here $\texttt{E} \{ \cdot \}$ is the expected value operator and the revenue rate $\pi > 0$ and the discount rate $\rho > 0$ are known constants.

The same additional assumption as in the deterministic case is added to the problem in the form of a state constraint. Again, assume that the firm has been accused of violating anti-trust laws in recent years and as such seeks to limit the market share to some percentage, $x_{\rm max} \in [0,1)$, to avoid breaking monopoly regulations. We will use stochastic optimal control to obtain the optimal advertising rate that maximizes Eq.~\ref{eq:obj3}, but it will also be monitored by an ASIF to intervene if the market share will exceed $x_{\rm max}$.

\subsubsection{Primary Controller}

The primary controller is derived using stochastic optimal control methods. It was shown in \cite{sethi_optimal_2018} that for the SDE given by 
% \begin{align}
%     \rm \boldsymbol{x}(t) = Q(\boldsymbol{x},\boldsymbol{u})\rm d t + B(\boldsymbol{x},\boldsymbol{u}) \rm d \boldsymbol{w}(t), ~ \boldsymbol{x}(0) = x_0,
% \end{align}
\begin{subequations}
    \begin{equation}
        {\rm d}\boldsymbol{x}(t) = Q(\boldsymbol{x},\boldsymbol{u})\rm d t + B(\boldsymbol{x},\boldsymbol{u}) \rm d \boldsymbol{w}(t),
    \end{equation}
    \begin{equation}
        \boldsymbol{x}(0) = x_0,
    \end{equation}
\end{subequations}
and associated objective function 
\begin{align}
    J = \texttt{E} \{ \int_0^{\infty} \phi(\boldsymbol{x},\boldsymbol{u})e^{-\xi t} dt \},
\end{align}
using It$\hat{\rm o}$'s Lemma, the Hamilton--Jacobi--Bellman (HJB) equation becomes 
% \begin{align}
% \begin{split}
%     \xi V(\boldsymbol{x}) = \max_{\boldsymbol{u}} \{ \phi(\boldsymbol{x},\boldsymbol{u}) + \frac{\partial V(\boldsymbol{x})}{\partial t}
%     + \frac{\partial V(\boldsymbol{x})}{\partial x}Q(\boldsymbol{x},\boldsymbol{u})  \\ + \frac{1}{2}\frac{\partial^2 V(\boldsymbol{x})}{\partial x^2}B(\boldsymbol{x},\boldsymbol{u})^2 
%     \}  
% \end{split}
% \end{align}
\begin{align} \label{eq:HJB_oa}
\begin{split}
    \xi V(\boldsymbol{x}) = \max_{\boldsymbol{u}} \{ \phi(\boldsymbol{x},\boldsymbol{u}) 
    + \frac{\partial V(\boldsymbol{x})}{\partial \boldsymbol{x}}Q(\boldsymbol{x},\boldsymbol{u}) \quad \quad \quad \\ + \frac{1}{2}\frac{\partial^2 V(\boldsymbol{x})}{\partial \boldsymbol{x}^2}B(\boldsymbol{x},\boldsymbol{u})^2 
    \},
\end{split}
\end{align}
where $V(\boldsymbol{x})$ is the current-valued value function. This is conveniently the form of the stochastic optimal advertising problem, and thus plugging \eqref{eq:sde_oa1} and \eqref{eq:obj3} into \eqref{eq:HJB_oa}, the HJB becomes
\begin{align} \label{eq:HJB_oa_pluggedin}
\begin{split}
    \rho V(\boldsymbol{x}) = \max_{\boldsymbol{u}}
    \{ 
    \pi \boldsymbol{x} - \boldsymbol{u}^2 + \frac{\partial V(\boldsymbol{x})}{\partial \boldsymbol{x}}(-\beta\boldsymbol{x} + r\boldsymbol{u}\sqrt{1 - \boldsymbol{x}}) \\ 
    + \frac{1}{2}\frac{\partial^2 V(\boldsymbol{x})}{\partial \boldsymbol{x}^2}(\sigma(\boldsymbol{x}))^2
    \}.
\end{split}
\end{align}
Taking the derivative of the bracketed terms on the right hand side of \eqref{eq:HJB_oa_pluggedin} with respect to $\boldsymbol{u}$ to find the critical point reveals that
\begin{align} \label{eq:optimal_u_intermed}
    \boldsymbol{u}^*(\boldsymbol{x}) = \frac{\frac{\partial V(\boldsymbol{x})}{\partial \boldsymbol{x}} r \sqrt{1 - \boldsymbol{x}}}{2}.
\end{align}
$\boldsymbol{u}^*(\boldsymbol{x})$ maximizes the expression in the brackets on the right hand side. Substituting \eqref{eq:optimal_u_intermed} into \eqref{eq:HJB_oa_pluggedin}, a solution of the current-valued value function is found to be
\begin{subequations}
    \begin{equation}
        V(\boldsymbol{x}) = \Bar{\lambda} \boldsymbol{x} + \frac{\Bar{\lambda}^2 r^2}{4 \rho},
    \end{equation}
    \begin{equation}
        \Bar{\lambda} = \frac{\sqrt{(\rho + \beta)^2 + r^2 \pi} - (\rho + \beta)}{r^2/2}.
    \end{equation}
\end{subequations}
Finally, the explicit formula for the optimal feedback control for the stochastic advertising problem is obtained as
\begin{align}
    \boldsymbol{u}^*(\boldsymbol{x}) = \frac{\Bar{\lambda}r \sqrt{1 - \boldsymbol{x}} }{2}.
\end{align}

\subsubsection{Stochastic Barrier Constraints}

The state constraint is identical to that of the deterministic optimal advertising problem. The safety constraint function is formulated as
\begin{align}\label{eq:h_soa}
    h_{soa}(\boldsymbol{x}) = -\boldsymbol{x}^2 + x_{\rm max}^2 \ge 0.
\end{align}
Using the SDE \eqref{eq:sde_oa1} and condition \eqref{eq:scbf_BC_NEW} in Thm.~\ref{thm:1}, the stochastic barrier constraint can be constructed for the ASIF-QP to use. With
\begin{align}
    \frac{\partial h_{soa}}{\partial \boldsymbol{x}} = -2\boldsymbol{x}, \quad \frac{\partial^2 h_{soa}}{\partial \boldsymbol{x}^2} = -2, 
\end{align}
we apply It$\rm \hat{o}$'s Lemma, obtaining
\begin{align}
    \begin{split}
    {\rm d}h_{soa}(\boldsymbol{x}) = \underbrace{(
    -2\boldsymbol{x}(-\beta \boldsymbol{x} + r\sqrt{1-\boldsymbol{x}}\boldsymbol{u})
    - (\sigma_a \boldsymbol{x})^2)
    }_{:=\Tilde{\mu}_{soa}}{\rm d}t \\ 
    + \underbrace{( -2\boldsymbol{x}(\sigma_{a}\boldsymbol{x})
    )}_{:=\Tilde{\sigma}_{soa}}{\rm d}\boldsymbol{w}.
    % \qquad \qquad \qquad \qquad \qquad
    \end{split}
\end{align}
Finally, the boundary constraints imposed on the control take the form
\begin{align} \label{eq:sbc_FINAL}
   \Tilde{\mu}_{soa} - \frac{\Tilde{\sigma}_{soa}^2}{h_{soa}(\boldsymbol{x})} \ge -h_{soa}(\boldsymbol{x})^2\alpha_2(h_{soa}(\boldsymbol{x})),
\end{align}
and are enforced by the QP. $\alpha_2$ is a strengthening function according to Definition~\ref{def:2}.

\subsection{Portfolio Optimization}

For the final application of CBFs to economics and finance, a classic portfolio optimization example is used. First introduced by Merton \cite{merton69}, the Merton problem is a problem in continuous-time finance wherein an investor seeks to maximize the expected discounted wealth by trading in a risky asset and risk-free bank account. We are interested in the state, $\boldsymbol{x} \in \mathbb{R}$, which is the investor's total invested wealth. The investor's total wealth is governed by the SDE
\begin{align} \label{eq:sde_po}
\begin{split}
    {\rm d} \boldsymbol{x}(t) = (\epsilon_{b} \boldsymbol{x}(t) + (\epsilon_{r} - \epsilon_{b})\boldsymbol{u}(t)){\rm d} t 
    + \sigma_{po}\boldsymbol{u}(t)\rm d \boldsymbol{w}(t),
\end{split}
\end{align}
where $\epsilon_b > 0$ is the continuously compounded rate of return of the risk-free bank account, and $\epsilon_r > 0$ represents the expected continuously compounded rate of growth of the risky asset. Both are assumed to be constant throughout the process. $\sigma_{po} > 0$ represents the volatility of the risky asset which is also constant throughout, and $\boldsymbol{u}(t)$ is the control variable which represents how much the investor has invested in the risky asset at time $t$ (with the remaining funds invested in the risk-free bank account). $\boldsymbol{w}(t)$ is a one dimensional Brownian motion.

The investor has a known finite investment horizon, meaning that the investor seeks to maximize their expected wealth over a certain time period, $t \in [0,T]$, where $T$ is the final time. Important to the problem is the investor's utility function, $U(\boldsymbol{x})$, which captures the preferences of the investor. Two common examples are the exponential utility function, and the hyperbolic absolute risk aversion (HARA) utility function \cite{cartea_algorithmic_2015}.
In this example, the exponential utility function will be used, which takes the form
\begin{align}
    U(\boldsymbol{x}) = -e^{-\gamma \boldsymbol{x}}, \quad \gamma > 0,
\end{align}
which is defined for all $\boldsymbol{x} \in \mathbb{R}$. The goal is thus to maximize the expected value of the exponential utility. In Chapter 5, Cartea et al. show that for this classic problem with an exponential utility function, the optimal amount to invest in the risky asset is a deterministic function of time \cite{cartea_algorithmic_2015}. They use the HJB equation as in Sec.~\ref{sec:soa} to obtain a control that maximizes the objective function. For brevity the derivation is omitted here. 

\subsubsection{Primary Controller}

The primary controller, $\boldsymbol{u}^*(t)$, is given by
\begin{align}\label{eq:optimal_po}
    \boldsymbol{u}^*(t) = \frac{\lambda}{\gamma \sigma_{po}} e^{-\epsilon_b(T-t)},
\end{align}
where $\lambda = \frac{\epsilon_r - \epsilon_b}{\sigma_{po}}$. $\lambda$ is known as the Sharpe ratio, or the market price of risk, and is a very widely used method for measuring risk-adjusted relative return. As the risk aversion, captured by $\gamma$, increases, the investment of wealth into the risky asset decreases, which is intuitive. 

\subsubsection{Stochastic Barrier Constraints}

The state constraint for the portfolio optimization problem takes a slightly different form than in the previous examples. Suppose that the investor must always have a certain amount of wealth, denoted $x_{\rm min}$, at any given time $t$. This is their emergency fund and it is of paramount importance to the investor that their wealth never go below this value under any circumstances. This can be formalized using the constraint function,
\begin{align}\label{eq:constraint_po}
    h_{po}(\boldsymbol{x}) = \boldsymbol{x}^2 - x_{\rm min}^2 \ge 0.
\end{align}
Furthermore, suppose that the investor occasionally takes a large portion of their total wealth and spends it, despite being told that this is not an optimal strategy by their financial advisor. In these cases, the investor never spends so much that their total wealth drops below $x_{\rm min}$, but it may come dangerously close. In these situations especially, an SCBF must be used in order to satisfy the constraint function \eqref{eq:constraint_po} for all time. Using the SDE \eqref{eq:sde_po} and condition \eqref{eq:scbf_BC_NEW}, the SCBF can be constructed as
\begin{align} \label{eq:sbc_PO}
   \Tilde{\mu}_{po} - \frac{\Tilde{\sigma}_{po}^2}{h_{po}(\boldsymbol{x})} \ge -h_{po}(\boldsymbol{x})^2\alpha_3(h_{po}(\boldsymbol{x})),
\end{align}
where $\alpha_3$ is another strengthening function, and
\begin{subequations}
    \begin{equation}
        \Tilde{\mu}_{po} = 2\boldsymbol{x}(\epsilon_{b} \boldsymbol{x} + (\epsilon_{r} - \epsilon_{b})\boldsymbol{u}) + (\sigma_{po}\boldsymbol{u})^2,
    \end{equation}
    \begin{equation}
        \Tilde{\sigma}_{po} = 2\boldsymbol{x}(\sigma_{po}\boldsymbol{u}).
    \end{equation}
\end{subequations}
The constraint is no longer suitable for a QP however. Since $\boldsymbol{u}$ appears nonlinearly in the constraint, a nonlinear program (NLP) must be used to satisfy \eqref{eq:sbc_PO} instead. 

% - explain what the optimization problem is
% - explain state constraints
% - derive primary controller
% - SBC constraints

% and therefore using the SDE given in Eq.~\ref{eq:dynamics1}, the barrier constraint to the QP is
% \begin{align}\label{eq:BC_advert}
%     BC(\boldsymbol{x}, \boldsymbol{u}) = 2\beta \boldsymbol{x}^2 - 2\boldsymbol{x}(1-\boldsymbol{x})\boldsymbol{u} + \alpha(h_{oa}(\boldsymbol{x})) \ge 0,
% \end{align}
% where $\alpha$ is a strengthening function as described in Sec.~\ref{sec:cbf}.

% \begin{enumerate}
%     \item Describe problem and cite literature
%     \item Derive optimal control solution using stochastic optimal control
%     \item Derive barrier constraints and have control barrier functions
%     \item Don't mention $\alpha$ yet
% \end{enumerate}

\section{Numerical Results}

The ASIF and optimal control solutions to each of the three economics problems were simulated in \textsc{Python} and the numerical results are summarized in this section. 

\subsection{Optimal Advertising}

\begin{figure} [ht!]
    \centering
    \vspace{-.2cm}
    % \centerline{\includesvg[inkscapelatex=false,width=1.05\columnwidth]{figs/archive/x_1.svg}}
    \centerline{\includegraphics[width=1.05\columnwidth]{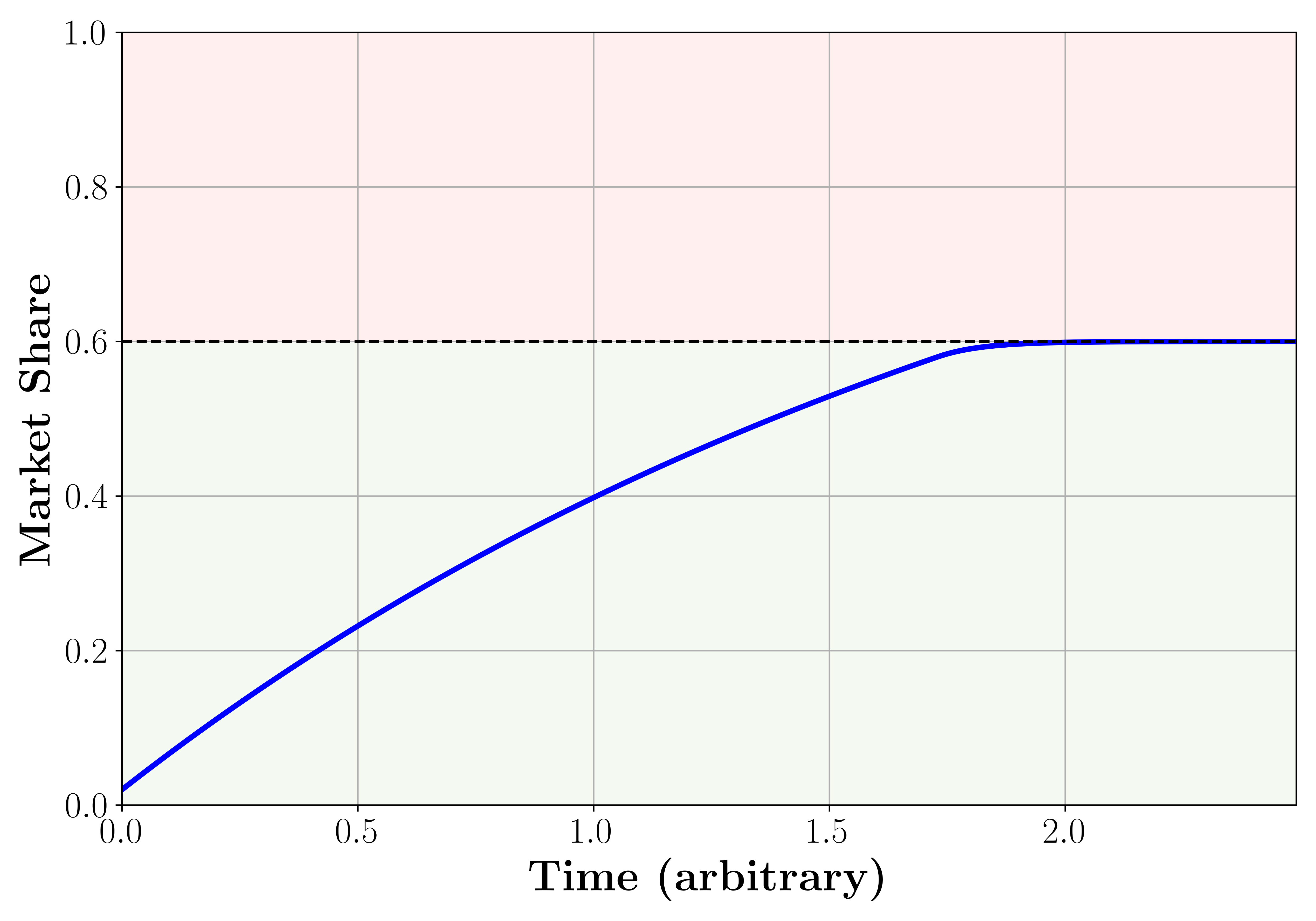}}
    \caption{Market share over time demonstrating compliance with anti-trust regulations for deterministic advertising.}
    \label{fig:x_1}
\end{figure}

Fig.~\ref{fig:x_1} plots the market share over time for the deterministic optimal advertising solution. The red region of the figure denotes the unsafe region which may result in anti-trust lawsuits and be detrimental to the firm's discounted profits. The green region denotes the safe portion of the state space. The firm began with a market share of 2\% and saw an increase as the advertising control $\boldsymbol{u}=a^{\kappa}$ was applied. The market share approached $x_{\rm max} = 0.6$ but never crossed the boundary, ensuring maximum profits for the firm while avoiding scrutiny from regulatory organizations. The strengthening function $\alpha$ used in the simulations was $\alpha(\cdot) = 10h_{oa}(\boldsymbol{x})$.

\begin{figure} [ht!]
    \centering
    \vspace{-.2cm}
    % \centerline{\includesvg[inkscapelatex=false,width=1.05\columnwidth]{figs/archive/u_1.svg}}
    \centerline{\includegraphics[width=1.05\columnwidth]{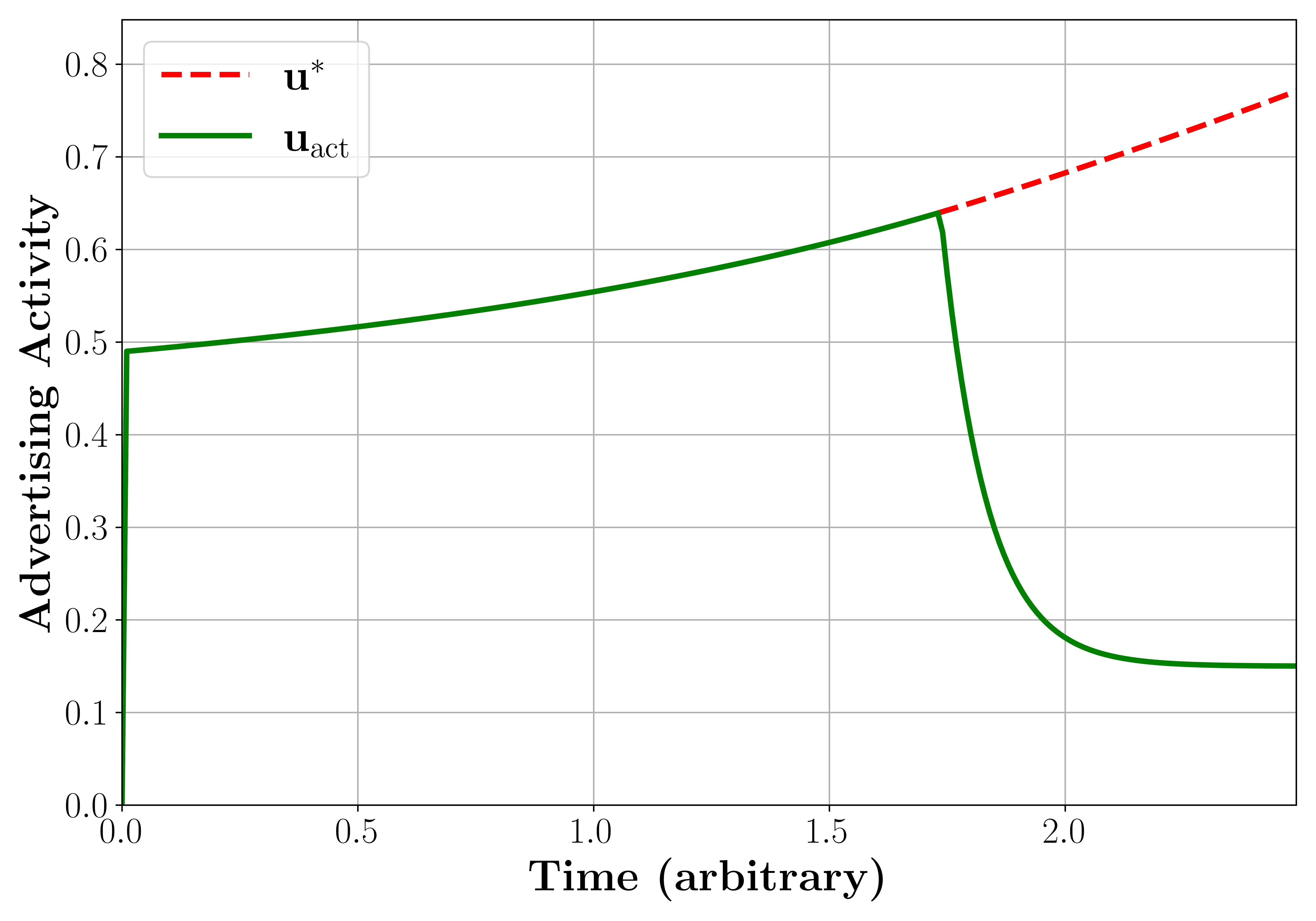}}
    \caption{Control inputs over time demonstrating safety interventions for deterministic advertising.}
    \label{fig:u_1}
\end{figure}

Fig.~\ref{fig:u_1} plots the advertising control over time where the optimal control is plotted with the red dotted line and the actual control used by the firm is plotted in green. The two were identical until the market share began to approach $x_{\rm max}$. The actual control used smoothly decreased from the optimal control, minimizing the distance between $\boldsymbol{u}^*$ and $\boldsymbol{u}_{\rm act}$ whilst satisfying the boundary constraints in Eq.~\ref{eq:BC_advert} and as a consequence, the state constraints in Eq.~\ref{eq:hoa}.

The average solver time was 0.0281 ms, and the maximum single solver time was 0.1202 ms. The QP was solved using \texttt{quadprog.solve\_qp}\cite{quadprog} running on a \textit{Dell XPS 15 9520 i9-12900HK} laptop. The simulation code is publicly available for all three problems\footnote{\texttt{https://github.com/davidvwijk/EconCBF}}. 

\subsection{Stochastic Optimal Advertising}

Due to the Brownian motion inherent to the stochastic optimal advertising problem, the results for a single random trial are shown, and then two Monte Carlo simulations using 1,000 samples each are analyzed. In the Monte Carlo simulations, the importance of the $\alpha_2$ strengthening function is explored.

\subsubsection{Single Trial}

\begin{figure} [ht!]
    \centering
    \vspace{-.2cm}
    % \centerline{\includesvg[inkscapelatex=false,width=1.05\columnwidth]{figs/archive/SCBF_x.svg}}
    \centerline{\includegraphics[width=1.05\columnwidth]{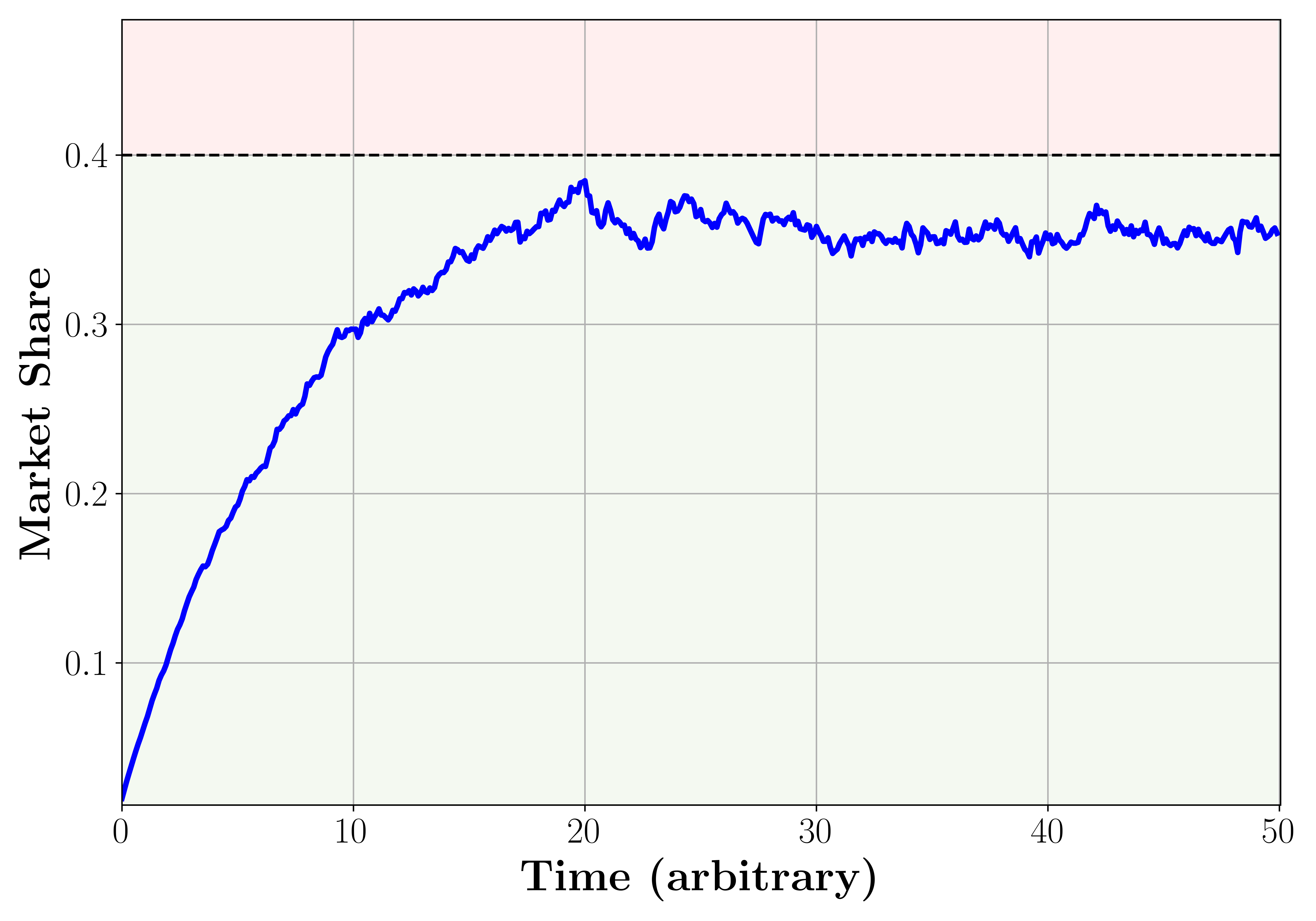}}
    \caption{Single trial market share over time demonstrating compliance with anti-trust regulations for stochastic advertising.}
    \label{fig:scbf_x}
\end{figure}

Fig.~\ref{fig:scbf_x} plots the market share of the firm over time for a single trial. As before, the red region of the figure denotes the unsafe region which may result in anti-trust lawsuits. The green region denotes the safe portion of the state space. The firm began with a market share of 2\% and saw an increase in market share as the optimal rate of advertising, $\boldsymbol{u}^*(\boldsymbol{x})$, was applied. Once the market share began approaching $x_{\rm max} = 0.4$, the optimal control was sporadically modified by the AISF-QP, keeping the stochastic system within the safe region, but maximizing the discounted profits whenever possible.

\begin{figure} [ht!]
    \centering
    \vspace{-.2cm}
    % \centerline{\includesvg[inkscapelatex=false,width=1.05\columnwidth]{figs/archive/SCBF_u.svg}}
    \centerline{\includegraphics[width=1.05\columnwidth]{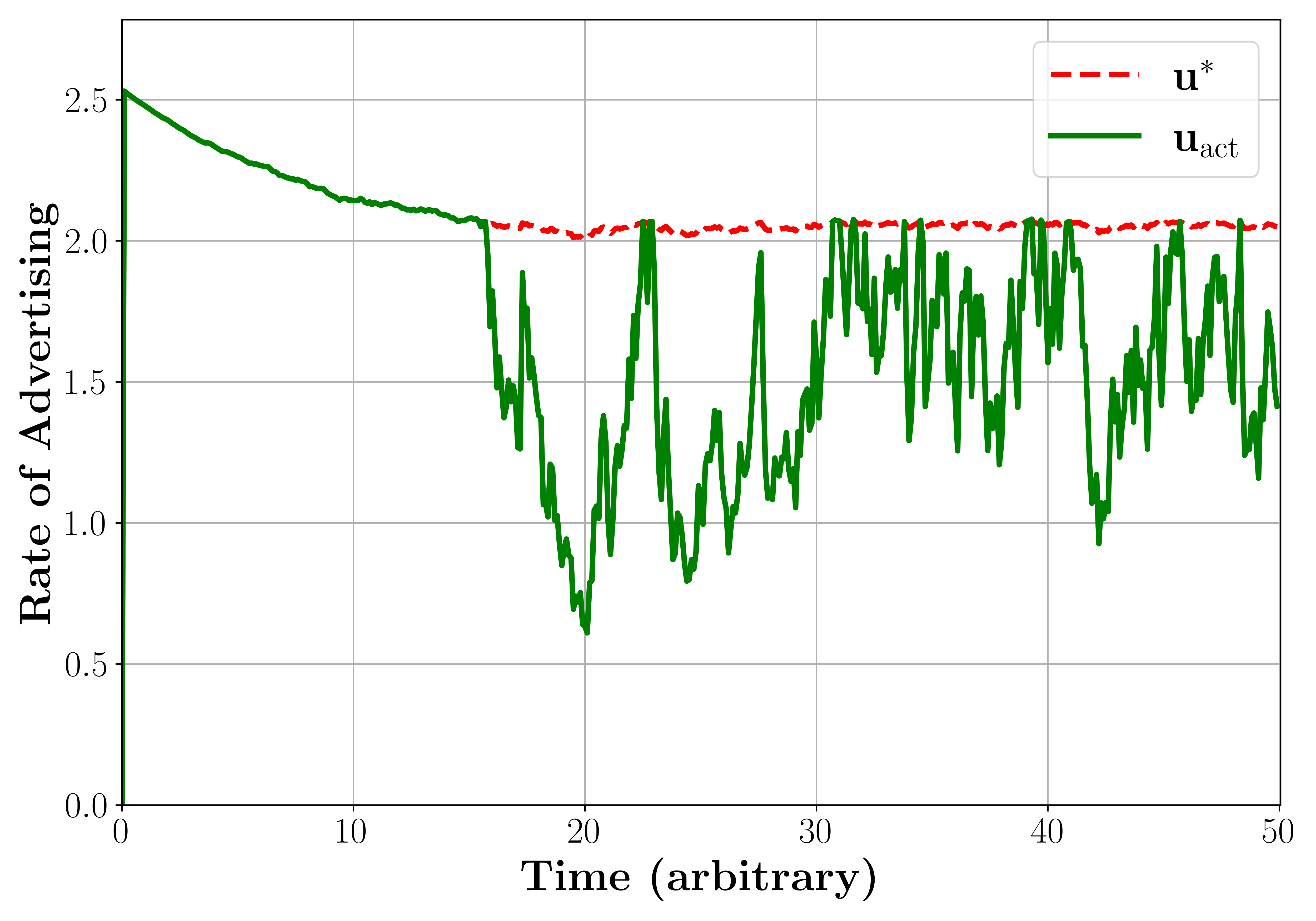}}
    \caption{Single trial control inputs over time demonstrating safety interventions for stochastic advertising.}
    \label{fig:scbf_u}
\end{figure}

This behavior can clearly be visualized in Fig.~\ref{fig:scbf_u} which plots the rate of advertising over time. The optimal rate of advertising is plotted red and the actual rate of advertising used by the firm is plotted in green. Since the stochastic barrier constraint accounts for the uncertainty in the dynamics, the market share did not graze the $x_{\rm max}$ line as in the deterministic example. For this single trial the $\alpha_2$ strengthening function is given by $\alpha_2(\cdot)= 500 h_{soa}(\boldsymbol{x})$.

The average solver time for the single trial was 0.0251 ms, and the maximum single solver time was 0.0635 ms, using the same machine as mentioned previously.

\subsubsection{Monte Carlo Simulations}

Two Monte Carlo (MC) simulations were performed with 1,000 samples each. The initial conditions were held fixed, along with the relevant constants described in Sec.~\ref{sec:soa}. Each MC simulation used a different strengthening function, $\alpha_2$, to demonstrate the important role that this function has on system behavior. The strengthening function bounds the rate of change of the barrier function, and in both cases took the form $\alpha_2(\cdot) = \eta h_{soa}(\boldsymbol{x})$ where $\eta$ is a positive constant which changes the slope of the bounding function. Higher values of $\eta$ result in less conservative barrier constraints.

\begin{figure} [ht!]
    \centering
    \vspace{-.2cm}
    % \centerline{\includesvg[inkscapelatex=false,width=1.05\columnwidth]{figs/archive/success_MC_1000.svg}}
    \centerline{\includegraphics[width=1.05\columnwidth]{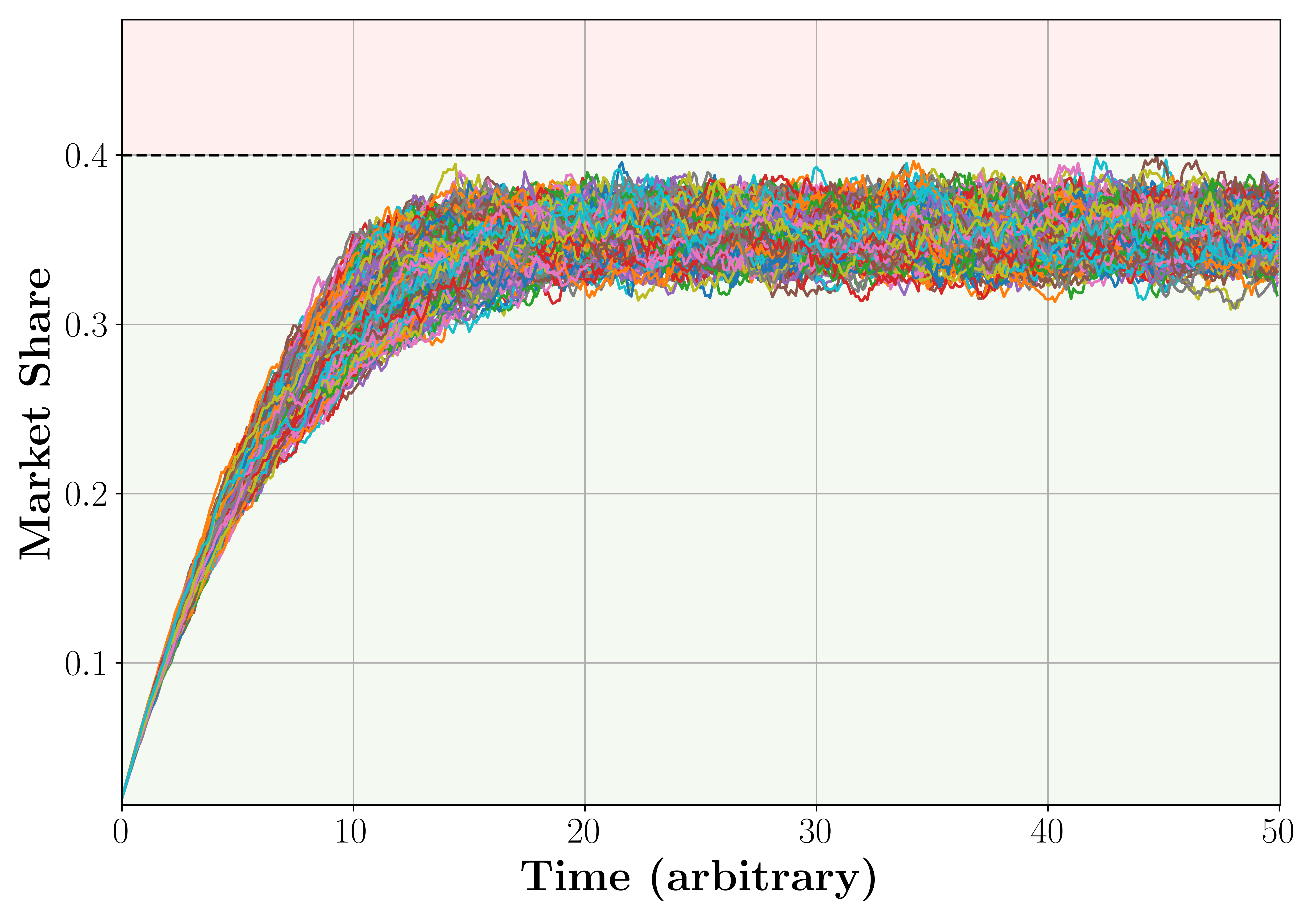}}
    \caption{Monte Carlo simulation for stochastic advertising with 1,000 samples, $\alpha_2(\cdot) = 500 h_{soa}(\boldsymbol{x})$.}
    \label{fig:mc1}
\end{figure}

Fig.~\ref{fig:mc1} plots market share over time for the 1,000 trials overlaid on the same graph. In this MC simulation, $\alpha_2(\cdot) = 500 h_{soa}(\boldsymbol{x})$ as with in the single trial results. Due to the nature of the diffusion coefficient function, $\sigma(\boldsymbol{x})$, the spread of trajectories began very tight, but widened as the market share increased. Once the market share approached $x_{\rm max}$, the curves began to stabilize, and as such, a near constant distribution of trajectories was obtained centered around $\boldsymbol{x} \approx 0.36$. During the 1,000 trials, there were no timesteps during which the market share exceeded $x_{\rm max}$. However, there were some instances that got very close to the boundary.

\begin{figure} [ht!]
    \centering
    \vspace{-.2cm}
    % \centerline{\includesvg[inkscapelatex=false,width=1.05\columnwidth]{figs/archive/success_MC_1000_alpha100.svg}}
    \centerline{\includegraphics[width=1.05\columnwidth]{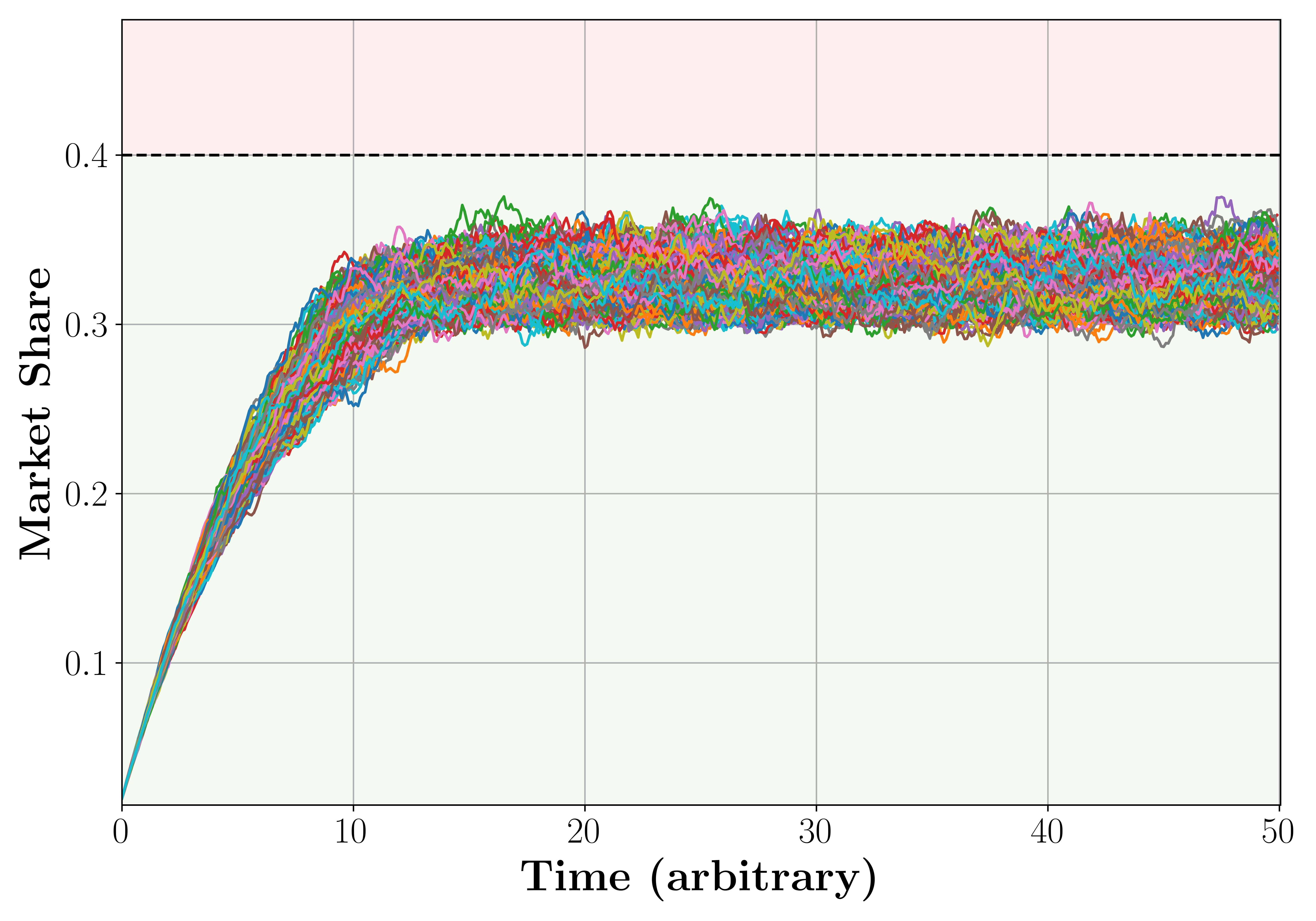}}
    \caption{Monte Carlo simulation for stochastic advertising with 1,000 samples, $\alpha_2(\cdot) = 100 h_{soa}(\boldsymbol{x})$.}
    \label{fig:mc2}
\end{figure}

Suppose the firm cannot afford even a single instance where the market share exceeds $x_{\rm max}$. In this case, the strengthening function shown in Fig.~\ref{fig:mc1} may not be appropriate. Fig.~\ref{fig:mc2} again plots market share over time for 1,000 trials overlaid on the same graph, but for a different strengthening function. In this MC simulation, $\alpha_2(\cdot) = 100 h_{soa}(\boldsymbol{x})$, making the strengthening function more conservative. As before, there was initially less noise in the market share, but as market share increased, so too did the associated noise. The samples again began to stabilize, but this time at around $\boldsymbol{x}\approx0.325$. Clearly, the strengthening function in this MC simulation was more conservative, reducing the likelihood of safety violations. The downside, however, was that the discounted profits obtained by a firm using this more conservative strengthening function would be lower than a firm using a less conservative strengthening function. 

\subsection{Portfolio Optimization}

For the portfolio optimization problem, the results for a single random trial are shown, and then a Monte Carlo simulation using 1,000 samples is analyzed.

In the numerical simulations, the rate of return of the risk-free bank account, $\epsilon_b$, was equal to 2\%. The expected rate of growth of the risky asset, $\epsilon_r$, was 16\%, and the volatility of the risky asset was quite high, at $\sigma_{po} = 11\%$. The initial investment was 550 thousand US dollars (USD), and the time horizon for the investor was 40 years. The investor used the optimal amount to invest in the risky asset from \eqref{eq:optimal_po}, but also used the condition in \eqref{eq:sbc_PO} to ensure that the portfolio always had at least 500 thousand USD in case of an emergency ($x_{\rm min}$ = 500). Assume that 15 years into the investment period, the investor made a very large purchase and the total wealth in the portfolio was reduced drastically. The purchase was such that the total wealth in the portfolio was 10\% above $x_{\rm min}$. For all simulations, the strengthening function $\alpha_3$ is given by $\alpha_3(\cdot) = h_{po}(\boldsymbol{x})/1{\rm e}16$.

\subsubsection{Single Trial}

\begin{figure} [ht!]
    \centering
    \vspace{-.2cm}
    % \centerline{\includesvg[inkscapelatex=false,width=1.05\columnwidth]{figs/archive/x_po.svg}}
    \centerline{\includegraphics[width=1.05\columnwidth]{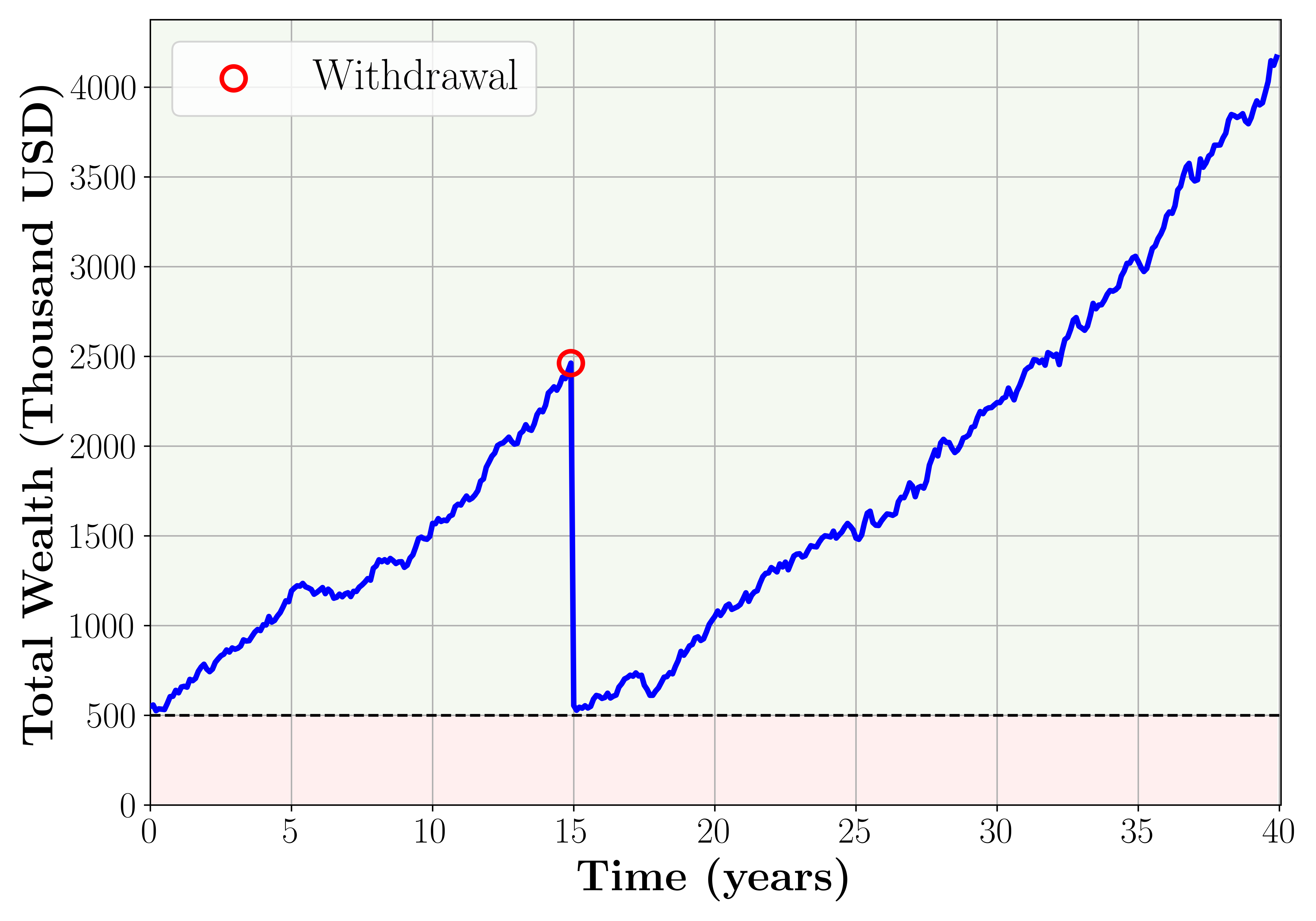}}
    \caption{Single trial of total wealth over time with large withdrawal demonstrating adherence to emergency fund criteria.}
    \label{fig:x_po}
\end{figure}

Fig.~\ref{fig:x_po} plots the total wealth of the investor over time. The red region of the plot depicts regions where the total wealth will dip below the wealth needed for an emergency, and the green region is the safe, or allowable, region. The initial total wealth was very close to $x_{\rm min}$, but once the wealth began compounding, there was a steady increase in total wealth. After 15 years, the investor withdrew a large sum of their wealth as specified above. After this large withdrawal, no further withdrawals occurred, and the total wealth reached over 4 million USD.

\begin{figure} [ht!]
    \centering
    \vspace{-.2cm}
    % \centerline{\includesvg[inkscapelatex=false,width=1.05\columnwidth]{figs/archive/u_po.svg}}
    \centerline{\includegraphics[width=1.05\columnwidth]{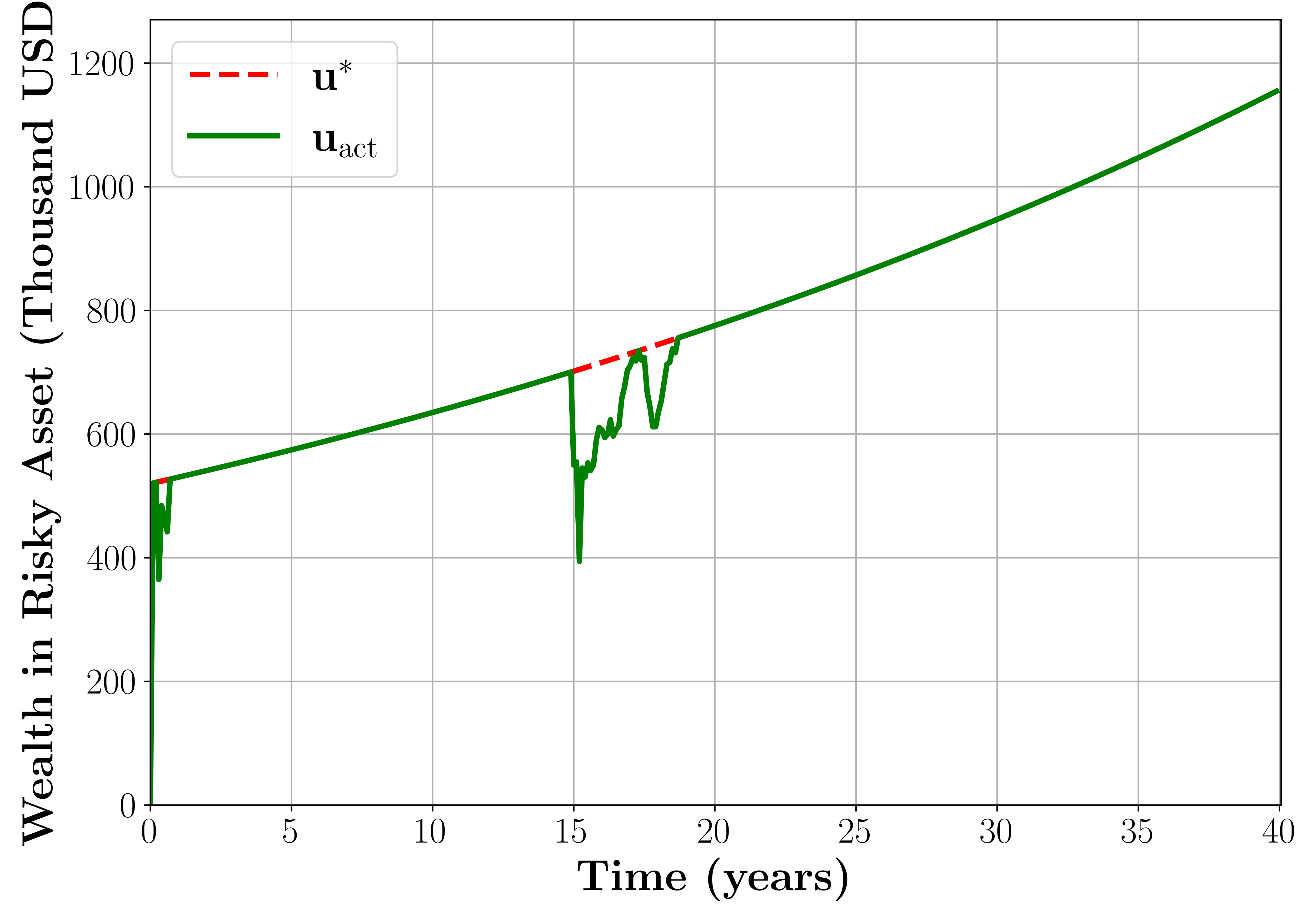}}
    \caption{Single trial control inputs over time demonstrating safety interventions for stochastic portfolio optimization.}
    \label{fig:u_po}
\end{figure}

Fig.~\ref{fig:u_po} plots the control used by the ASIF-NLP, which is the wealth invested in the risky asset over time. As usual, the optimal control is plotted in red, and the control actually used by the investor is plotted in green. Initially, since the total wealth began near the minimum wealth threshold, there was a large deviation in the optimal control and the actual control used. The amount of wealth actually invested in the risky asset was smaller than the desired amount due to the volatility of the asset. Thus, the constraints on the ASIF-NLP ensured that the total wealth did not dip into the unsafe region. When the major withdrawal occurred at $t=15$ years, again there was a deviation between the desired control and the actual control used. Since the withdrawal put the total wealth very near $x_{\rm min}$, the amount invested in the risky asset was decreased and the majority of the wealth was placed into the safe, risk-free bank. The wealth invested in the risky asset remained low during this period, and thus the total wealth increased rather slowly. However, once the total wealth of the investor was sufficiently far from $x_{\rm min}$, $\boldsymbol{u}^*$ was used for the remainder of the investment period. The average solver time for the single trial was 0.683 ms, and the maximum single solver time was 1.717 ms.

\subsubsection{Monte Carlo Simulation}

\begin{figure} [ht!]
    \centering
    \vspace{-.2cm}
    % \centerline{\includesvg[inkscapelatex=false,width=1.05\columnwidth]{figs/archive/mc_1000_po.svg}}
    \centerline{\includegraphics[width=1.05\columnwidth]{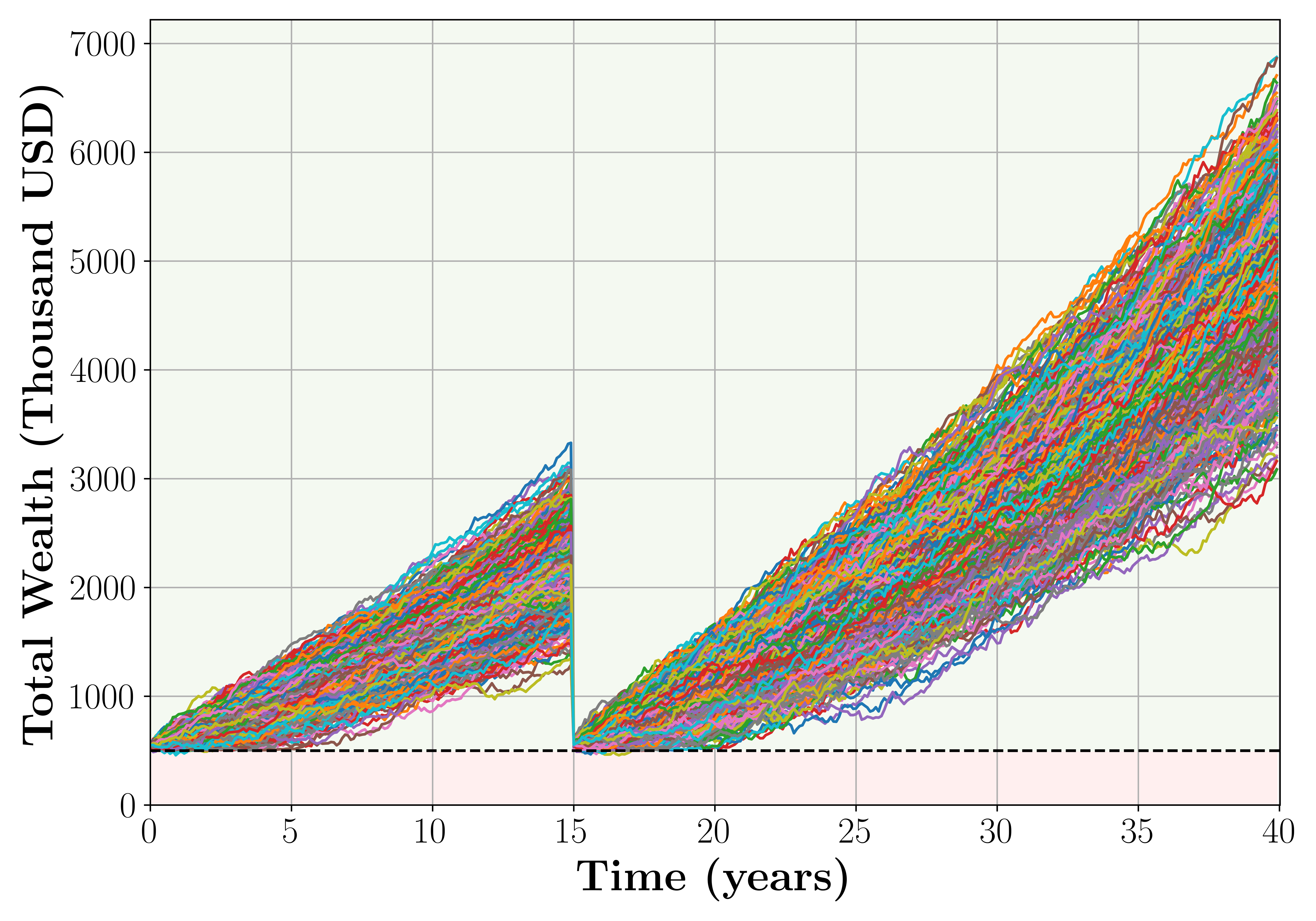}}
    \caption{1,000 sample Monte Carlo simulation for portfolio optimization with large withdrawal.}
    \label{fig:mc_po}
\end{figure}

A Monte Carlo simulation was performed using 1,000 samples. The initial conditions, strengthening function, withdrawal specification, time horizon and problem constants were all held fixed, and were as described for the single trial. The difference in potential outcome resulted solely from the stochasticity due to the risky asset. Fig.~\ref{fig:mc_po} plots the total wealth of 1,000 simulated investors over the 40 year time horizon. The total wealth remained within the safe region for 99.978\% of the timesteps (one timestep being 0.1 years). However, due to two main factors, there were moments where the total wealth dipped slightly into the unsafe region, below $x_{\rm min}$. The first factor was that the initial investment was too close to the boundary of the safe region. Occasionally it was possible that the ASIF-NLP would not be able to find a solution to satisfy the constraint on control. Secondly, much in the same vein, the withdrawal was too large in some scenarios. In these cases, the large withdrawal could cause slight safety violations. Thus these violations were due not to the ASIF-NLP and the condition given by \eqref{eq:sbc_PO}.

\section{Conclusion}

This paper applies control barrier functions to a class of dynamical systems derived from economics and finance which are fields not often considered in safety-critical control. An optimal advertising example is used, where the firm must maximize discounted profit but must adhere to constraints on the market share. An optimal control solution is derived, as well as a CBF for constraint enforcement. Simulation results demonstrate the effectiveness and computational efficiency of the proposed solution. A more complex example of stochastic optimal advertising is studied, which more closely reflects real economic systems due to the stochasticity of the problem. To handle the uncertainties in the dynamics governing the market share, an optimal solution is obtained using stochastic optimal control, and stochastic CBFs (SCBFs) are developed and implemented in a numerical simulation. These SCBFs enforce the same constraints posed in the deterministic advertising example. The effect of the strengthening function is explored using two Monte Carlo simulations. Lastly, a portfolio optimization problem with a volatile asset is considered, and SCBFs are used to enforce a constraint on the minimum total wealth of the investor. Future work may consider more complex, high-dimensional portfolio management problems where slack variables on SCBFs may need to be introduced.

% % \begin{enumerate}
% %     \item Summarize what we did
% %     \item Explain why its important and when its useful
% %     \item Refer to numerical study
% %     \item Extensions for future work
% % \end{enumerate}

% % -reducing math burden
% % -non-affine, relative degree issue
% % -not perfect but serves purpose in some cases

% \section*{Acknowledgment}

\bibliographystyle{ieeetr}
\bibliography{references}

\end{document}